\documentclass[twocolumn,prd,aps,superscriptaddress,showpacs,
nofootinbib,preprintnumbers,floats,floatfix]{revtex4}

\usepackage{amsmath}
\usepackage{amssymb}
\usepackage{latexsym}
\usepackage{graphicx}
\usepackage{hyperref}
\usepackage{bm}
\usepackage{color}
\usepackage{ifthen}

\newcommand{\fnl}{\ensuremath{f_{\mathrm{NL}}}}
\newcommand{\taunl}{\ensuremath{\tau_{\mathrm{NL}}}}
\newcommand{\gnl}{\ensuremath{g_{\mathrm{NL}}}}
\newcommand{\Nf}{N_{\mathrm{f}}}
\newcommand{\Nh}{\bar{N}}
\newcommand{\Mp}{M_{\mathrm{P}}}
\newcommand{\Ps}{\mathcal{P}}
\newcommand{\Neff}{N_{\mathrm{eff}}}
\newcommand{\Or}{\mathrm{O}}
\renewcommand{\d}{\mathrm{d}}

\newcommand{\alternative}[2]{#1}

\newcommand{\para}[1]{\par\vspace{1mm}\noindent\textbf{{#1}.}}

\newcommand{\arxiv}[1]{\href{http://arxiv.org/abs/#1}{#1}}
\newcommand{\arxivnew}[1]{\href{http://arxiv.org/abs/#1}{arXiv:#1}}

\newcommand{\etal}{et al.}

\newboolean{comments}
\setboolean{comments}{true}
\newcommand{\comment}[2]{\ifthenelse{\boolean{comments}}{\textcolor{red}{[\textsf{\textbf{{#1}}}: }\textcolor{blue}{\textsf{{#2}}}\textcolor{red}{]}}{}}

\begin{document}

\title{Non-gaussianity in axion N-flation models: detailed predictions
and mass spectra}

\author{Soo A~Kim} 
\affiliation{Asia Pacific Center for Theoretical Physics, Pohang,
  Gyeongbuk 790-784, South Korea}    
\author{Andrew R.~Liddle} 
\affiliation{Astronomy Centre, University of Sussex,
  Brighton BN1 9QH, United Kingdom} 
\author{David Seery}
\affiliation{Astronomy Centre, University of Sussex, Brighton BN1 9QH,
  United Kingdom} 
\date{\today}

\begin{abstract}
We have recently shown \cite{KLS} that multi-field axion N-flation can
lead to observable non-gaussianity in much of its parameter range,
with the assisted inflation mechanism ensuring that the density
perturbations are sufficiently close to scale invariance. In this
paper we extend our analysis in several directions. In the case of
equal-mass axions, we compute the probability distributions of
observables and their correlations across the parameter space. We
examine the case of unequal masses, and show that the mass spectrum
must be very densely packed if the model is to remain in agreement
with observations. The model makes specific testable predictions for
all major perturbative observables, namely the spectral index,
tensor-to-scalar ratio, bispectrum, and trispectrum.
\end{abstract}


\maketitle

%
%
\section{Introduction}

In a recent paper \cite{KLS} we identified a new mechanism for
generating an observably large non-gaussianities during inflation.
The effect is due to diverging trajectories near a maximum of the
potential, where---in a single-field model---the field would have a
large effective mass in Hubble units.  Such large masses are excluded
in the single-field case because the scalar spectral index is too far
from unity, but in the multi-field case the assisted inflation
phenomenon \cite{LMS} generates a spectrum that can be compatible with
current observational constraints \cite{WMAP7} without simultaneous
suppression of the dimensionless bi- or tri-spectra.  As a specific
example, we implemented this mechanism in the multi-field axion
N-flation model~\cite{DKMW}. An alternative implementation of the mechanism in a hybrid inflation context was recently given by Mulryne et al.~\cite{MOR}.

The purpose of the present paper is to provide a more detailed
phenomenological description of the axion \mbox{N-flation} model. In the case where all
fields have equal mass, which was assumed throughout Ref.~\cite{KLS},
we provide a comprehensive analysis of all key observables, analyzing
the probability distributions inherited from randomness in the initial
conditions, their dependence on model parameters, and their
intercorrelations.  We extend the analysis to the case where the
fields are distributed with a spectrum of masses, and obtain tight
constraints on the packing of the mass fraction necessary to maintain
agreement with observations.

%
%
\section{The axion N-flation model}

The axion N-flation model is based on a set of $\Nf$ uncoupled fields,
labelled $\phi_i$, each with a potential \cite{DKMW}
\begin{equation}
	V_i = \Lambda_i^4 \left ( 1- \cos \alpha_i \right) \,,
\end{equation}
where $\alpha_i = 2\pi\phi_i/f_i$ and $f_i$ is the $i^{\mathrm{th}}$
axion decay constant.  More generally, couplings may exist between the
fields but we will not consider these.  The mass of each field in the
minimum of the potential satisfies $m_i = 2\pi \Lambda_i^2/f_i$, and
the angular field variables $\alpha_i$ lie in the range
$(-\pi,+\pi]$. Without loss of generality we will set initial
  conditions with all $\alpha_i$ positive.  If only a single field is
  present this model is known as natural inflation~\cite{natural}.

One motivation for N-flation was to avoid the requirement for
super-Planckian field values \cite{DKMW}, which are invoked in many
single-field models. If one literally imposes $|\phi| < \Mp$ (where
$\Mp \equiv (8 \pi G)^{-1/2}$ is the reduced Planck mass) this
requires $f_i<2\Mp$ for each $i$. However, it would be reasonable to
regard this condition as a guideline rather than mandatory.

\subsection{The amount of inflation}

Any inflationary model must provide sufficient \mbox{$e$-foldings} to resolve
the classical cosmological problems. For a given set of initial angles
$\alpha_i^\ast$ one finds
\begin{eqnarray}
N_{\mathrm{tot}} 
 & \simeq & - \left( \frac{f_i}{2\pi \Mp} \right)^2 
   \int_{\alpha_i^\ast}^{\alpha_{\rm end}} 
   \sum_i \frac{V_i\, \d\alpha_i}{\partial V_{i} / \partial \alpha_i} 
			\nonumber \\
 & \simeq& \sum_i
\left(\frac{f_i}{2\pi \Mp}\right)^2 \ln \frac{2}{1+\cos \alpha_i^\ast}
\,, 
\label{eq:ntot}
\end{eqnarray}
where in the second line we have ignored a small correction from the
location of the end of inflation.  The sum is dominated by fields
whose initial angle is close to $\pi$.  However, for any reasonable
distribution of $\alpha_i^\ast$, the logarithm means that many fields
must cooperate to yield sufficient $e$-foldings---unless $f$ is
extremely large in Planck units, as in natural inflation. As remarked
above, this conflicts with the goal of maintaining sub-Planckian field
excursions.  Our successful models typically feature hundreds or
thousands of fields.

If the initial conditions are taken to be distributed uniformly in
angle, one can average to find
\begin{eqnarray}
\left\langle N_{\rm tot} \right\rangle 
  &\simeq& \frac{1}{4\pi^2 \Mp^2}\frac{\sum_i f_i^2 }{\pi}
		 \int^\pi_0 \d\alpha_{i}
		 \left[\ln \frac{2}{1+\cos\alpha_i}\right] \,,\nonumber\\
  &\simeq& \frac{ \ln 2}{2\pi^2} \frac{\sum_i f_i^2}{\Mp^2} \, ,
\end{eqnarray}
which confirms the requirement for many fields.

Having obtained sufficient inflation, we must identify the epoch at
which observable perturbations were generated. This requires knowledge
of the entire history of the Universe, including reheating, and is
therefore subject to some uncertainty.  We follow Ref.~\cite{Nstar} to
obtain the time at which the present horizon scale $k=a_0H_0$ crossed
outside the horizon, expressed in $e$-foldings before the end of
inflation:
\begin{equation}
N_{\rm hor}
 \approx 67  + \frac{1}{4} \ln \frac{V_{\rm hor}^2}{\rho_{\rm end} \Mp^4}
                   + \frac{1}{12} \ln \frac{\rho_{\rm reh}}{\rho_{\rm end}} \, .
\end{equation}
The last term might typically be $-5$ \cite{Nstar}. Here `hor', `reh'
and `end' denote values at horizon crossing and at the end of
reheating and inflation, respectively. To evaluate these terms, we
must ensure that models are normalized to reproduce the correct
amplitude of density perturbations.

For a single field, the appropriate $N_{\rm hor}$ depends how close
the initial position lies to the maximum; we find
\begin{eqnarray}
N_{\rm hor}
 &\simeq& 54 \quad (\mbox{near the maximum}) \,, \nonumber \\ 
 &\simeq& 59 \quad (\mbox{away from the maximum}) \,.
\end{eqnarray}
Near the hilltop, we require fewer $e$-folds because the flatness of
the potential implies we require smaller $H_\ast$. Therefore, the
duration of the radiation era which follows inflation is shorter.

We aim to verify that $N_{\rm hor}$ does not shift significantly when
$\Nf \gg 1$.  Taking $\Nf$ of order $10^3$, we find
\begin{equation}
N_{\rm hor} \simeq 57 \,.
\end{equation}
We have assumed $f =\Mp$ but the result is relatively insensitive to
this choice.  We conclude there is no significant change to $N_{\rm
  hor}$.

Finally, we note that the constraints we impose are not evaluated at
the present-day horizon scale 
\begin{equation}
k_{\rm hor} =a_0 H_0 = \frac{h}{3000} \, {\rm Mpc}^{-1}
\approx 0.00023 \, {\rm Mpc}^{-1} \,,
\end{equation}
but rather at $k_{*} = 0.002 \,
        {\rm Mpc}^{-1}$.  This scale is inside the present horizon,
        with $N_{*} \simeq N_{\rm hor}-2$. Overall, we conclude that
        in the multi-field case it remains a reasonable hypothesis
        that the pivot scale crossed the horizon around 50 to 60
        $e$-foldings before the end of inflation.  In what follows we
        will consider only these two values.

\subsection{Perturbations}

We calculate observables using the $\delta N$ formula \cite{deltaN},
which measures fluctuations in the total
$e$-foldings of expansion owing to field perturbations.  We define $\epsilon$-like slow-roll
parameters for each field,
\begin{equation}
	\epsilon_i \equiv \frac{\Mp^2}{2}
	\left( \frac{V_i'}{V_i} \right)^2\,,
	\label{eq:epsi}
\end{equation}
where a prime denotes a derivative with respect to $\phi_i$, and no
summation over $i$ is implied.  The global slow-roll parameter
$\epsilon \equiv -\dot{H}/H^2$ can be written as a weighted sum
$\epsilon \simeq \sum_i (V_i/V)^2 \epsilon_i$, in which each field
contributes according to its share of the total energy density. We
must have $\epsilon < 1$ during inflation.

We work in the horizon-crossing approximation, in which the dominant
contribution to each observable is assumed to arise from fluctuations
present only a few $e$-folds after horizon exit of the wavenumber
under discussion.  After smoothing the universe on a superhorizon
scale somewhat smaller than any scale of interest, the
horizon-crossing approximation becomes valid whenever the ensemble of
trajectories followed by smoothed patches of the universe approaches
an attractor.  The validity of the horizon-crossing approximation was
recently discussed by Elliston {\etal}~\cite{HCA}, the expectation
being that it is a good approximation in our case. Numerical
calculations supporting this conclusion were reported in
Ref.~\cite{MSW}.

The observables of interest, defined in the conventional way
\cite{PDP}, are the scalar spectral index $n$, the tensor-to-scalar
ratio $r$, the bispectrum $\fnl$, and the trispectrum parameters
$\taunl$ and $\gnl$. They are given by
\begin{eqnarray}
\Ps_\zeta & = & \frac{H_*^2}{4\pi^2} \sum_i N_{,i}N_{,i} =
\frac{H_*^2}{8\pi^2 \Mp^2} \sum_i \frac{1}{\epsilon_i^*}\,;
\label{eq:P} \\
n-1 & = &-2\epsilon_* - \frac{8\pi^2}{3H_*^2} \sum_j
\frac{\Lambda_j^4}{f_j^2}  
    \frac{1}{\epsilon_j^*} \Big/ \sum_i
    \frac{1}{\epsilon_i^*} \,; \label{eq:n} \\  
r & = & \frac{2}{\pi^2 \Ps_\zeta} \frac{H_*^2}{\Mp^2} = 16 \Big/ \sum_i
\frac{1}{\epsilon_i^*}\,; \label{eq:r} \\
\frac{6}{5} \fnl & = & \nonumber
\frac{\sum_{ij}N_{,i}N_{,j}N_{,ij}}{\left(\sum_k N_{,k}N_{,k}\right)^2} \\
& = & \frac{r^2}{128} \sum_i \frac{1}{\epsilon_i^*} \frac{1}{1+\cos
  \alpha_i^*} \,;\quad
  \label{eq:fnl}\\
\taunl \nonumber
  & = & \frac{\sum_{ijl} N_{,i}N_{,j}N_{,il}N_{,jl}}{(\sum_k N_{,k} N_{,k})^3} \\
  &=& \frac{r^3}{512}\sum_i
\frac{1}{2\epsilon_i^\ast}\frac{1}{(1+\cos\alpha_i^*)^2} \,;
    \label{eq:taunl}\\
\frac{54}{25} \gnl \nonumber
  & = & \frac{\sum_{ijl} N_{,i}N_{,j}N_{,l}N_{,ijl}}{(\sum_k N_{,k} N_{,k})^3}\\
  &=& \frac{r^3}{512}\sum_i
\frac{1}{2\epsilon_i^\ast}\frac{1-\cos\alpha_i^*}{(1+\cos\alpha_i^*)^2} \,,
      \label{eq:gnl}
\end{eqnarray}
where $N_{,i}$, $N_{,ij}$ and $N_{,ijk}$ are respectively the first,
second and third derivatives of $N$ with respect to field values at
time $\ast$, corresponding to evaluation at the pivot scale determined
as in Eq.~(\ref{eq:ntot}).  In writing
Eqs.~\eqref{eq:fnl}--\eqref{eq:gnl}, any intrinsic non-gaussianity
among the field perturbations at horizon crossing has been neglected.
This is a good approximation whenever the bi- and tri-spectrum
parameters are large enough to be observable
\cite{Maldacena,bisp,VW,trisp}.  Our sign convention for $\fnl$
matches that used in WMAP papers~\cite{WMAP7}, and the non-gaussianity
is predicted to be of local type. The observed amplitude of
perturbations is obtained by adjusting the $\Lambda_i$ to give an
appropriate value of $H_*$.

Under a quadratic approximation to each potential, it can be shown
that Eqs.~\eqref{eq:r} and \eqref{eq:fnl} recover their single-field
values of order $\sim 1/N_\ast$ \cite{single,VW}, making $\fnl$
undetectably small.  The spectral index can be shown to be less than
its single-field value $1-2/N_*$ \cite{LR} with equality only in the
equal-mass case. Its value for a given choice of parameters must be
computed numerically \cite{KLs}.  However these results change
whenever the initial conditions populate the hilltop region.

\section{The equal-mass case}

In Ref.~\cite{KLS} we considered only the case where all fields have
the same potential. In this section we make the same assumption, but
carry out a much more detailed analysis of the phenomenology.  The
scale $\Lambda \equiv \Lambda_i$ is fixed from the observed amplitude
of $\Ps_\zeta$, leaving $f \equiv f_i$ and $N_{\rm f}$ as adjustable
parameters.

The initial conditions are drawn randomly from a uniform distribution
of angles $\alpha_i$, with several realizations to explore the
probabilistic spread. This choice seems plausible in light of the
approximate circular symmetry of axion potentials. One could of course
envisage other probability distributions, but to obtain a successful
model there must be some reasonable probability of populating the
hilltop region. Since those fields dominate the statistics of the
density perturbations, any probability distribution which is
approximately flat near the hilltop can be expected to yield similar
results.

\subsection{Analytic approximations}

The $\epsilon_i$ approach zero for fields close to the hilltop, so
each summation in Eqs.~\eqref{eq:P}--\eqref{eq:gnl} is dominated by
those fields with the smallest $\epsilon_i$.  Suppose some number
$\Nh$ of such fields have roughly comparable $\epsilon_i$, of order
$\bar{\epsilon}$.

\para{Quantum diffusion} Near the hilltop, the parameters $\epsilon_i
\sim \bar{\epsilon}$ are small and the classical motion of each axion
becomes small.  In this region, two new effects emerge.  First,
for sufficiently small $\bar{\epsilon}$ the classical motion of
individual fields can be dominated by quantum fluctuations, but
typically this is not of concern unless the fields involved contribute
non-negligibly to the energy density.  In this case one can expect
density fluctuations of order unity, leading to a phase of
`topological inflation' \cite{topo}.  Second, the hilltop is a
singularity of the e-folding history, $N$, as a function of the
initial field values.  If these initial values are chosen too close to
the singularity then the Taylor expansion used to obtain
Eqs.~\eqref{eq:P}--\eqref{eq:gnl} becomes unreliable.

We consider the constraints in turn, beginning with the issue of
singularities in $N$. These emerge from Eq.~\eqref{eq:ntot} in the
limit $\alpha_i^\ast \rightarrow \pi$.  By repeated differentiation,
we conclude that the Taylor expansion is trustworthy unless
\begin{equation}
	\delta_i^\ast \equiv | \alpha_i^\ast - \pi | \lesssim
		\frac{|\delta \phi_\ast|}{f_i} 
	\sim \frac{H_\ast}{\Mp} ,
	\label{eq:delta-n-breakdown}
\end{equation}
where the final approximate equality applies in a model for which $f
\sim \Mp$, and the field fluctuation $\delta \phi$ should be estimated
at the time of horizon exit, which we continue to label
$\ast$. Assuming $\Nh$ axions dominate the spectrum with comparable
$\delta_i^\ast \sim \delta_\ast$ and $f_i \sim f$, the observed
amplitude of density fluctuations $\Ps_{\zeta} \simeq 2 \times 10^{-9}$ requires
\begin{equation}
	\delta_\ast \approx \Nh^{1/2}
		\frac{f}{\Mp^2} \frac{|\delta\phi_\ast|}{\Ps_\zeta^{1/2}}
	\sim
		\Nh^{1/2} \frac{H_\ast}{\Mp} \Ps_{\zeta}^{-1/2} .
	\label{eq:density-normalization}
\end{equation}
For $\Nh = \Or(10)$,
Eqs.~\eqref{eq:delta-n-breakdown}--\eqref{eq:density-normalization}
imply that a breakdown of the perturbative $\delta N$ formula cannot
occur unless at least one $f_i$ is a few orders of magnitude less than
the Planck scale.  For $\Nh \gg 1$ a more extreme tuning of some $f_i$
is required.  As explained above, after drawing initial conditions
$\alpha_i^\ast$ within our numerical simulations, we adjust the Hubble
scale $H_\ast$ to satisfy Eq.~\eqref{eq:density-normalization} by a
suitable normalization of the scales $\Lambda_i$.

Under normal circumstances the power spectrum is monotonically
increasing with time and therefore
Eq.~\eqref{eq:density-normalization} guarantees that the adiabatic
trajectory is stable to quantum fluctuations.  But it should also be
checked that when the final axion rolls to its minimum, forcing the
correlation functions of $\zeta$ to their horizon-crossing values
Eqs.~\eqref{eq:P}--\eqref{eq:fnl}, its fluctuations are not large
enough to initiate an unwanted phase of topological
inflation. Therefore we require
\begin{equation}
	\delta_{\mathrm{roll}} \gtrsim \frac{H_{\mathrm{roll}}^3}{\Lambda^3}
		\frac{f}{\Lambda}
		\sim \frac{\Lambda^2}{\Mp^2} ,
	\label{eq:roll-diffusion}
\end{equation}
where in the final step we have taken $f \sim \Mp$ and, because the
final axion field dominates the potential by definition, we have
estimated $H_{\mathrm{roll}} \sim \Lambda^2 / \Mp$.

Our numerical simulations do not take quantum diffusion into account,
so we will usually wish to impose the stronger requirement that
diffusion does not occur for any field.  This also avoids the
possibility that the final axion field diffuses to sufficiently small
values that Eq.~\eqref{eq:roll-diffusion} is violated.  It is
sufficient to demand that Eq.~\eqref{eq:density-normalization} bounds
$\delta$ away from the quantum diffusion regime.  Taking $f \sim \Mp$
this requires
\begin{equation}
	H_\ast^2 \Mp^2 \lesssim \Nh^{1/2} \Lambda^4 \Ps_{\zeta}^{-1/2} .
	\label{eq:no-diffusion}
\end{equation}
Once the spectrum has been correctly normalized,
Eq.~\eqref{eq:no-diffusion} can be interpreted as a bound on the
number of axions, $\Neff$ which contribute an energy density of order
$\Lambda^4$,
\begin{equation}
	\Neff \lesssim \Nh^{1/2} \Ps_\zeta^{-1/2} .
	\label{eq:quantum-diffusion-bound}
\end{equation}
Note that $\Neff > \Nh$, since an axion which contributes an energy
density of order $\Lambda^4$ will not contribute to $\Ps_\zeta$ unless
its contribution is enhanced by proximity to the hilltop.  A similar
discussion was given by Huang \cite{Huang}.

\para{Observable quantities} We now proceed to study the various
observable quantities, each of which has a different scaling with
$\Nh$. The spectrum, $\Ps_\zeta$, scales like $\Nh$ copies of a
single-field model with slow-roll parameter $\bar{\epsilon}$, whereas
$r$ is \emph{reduced} by a factor $\Nh$ compared to its value in the
same single-field model.  The spectral index can be written
\begin{equation}
	n - 1 \approx - 2 \epsilon_\ast - 8 \pi^2
		\left( \frac{\Mp}{f} \right)^2 \Big/
		\sum_i (1 - \cos \alpha_i^\ast) \,,
	\label{eq:napprox}
\end{equation}
and is independent of $\Nh$.  Instead, the summation in the
denominator receives contributions from all fields.  Assuming $f$ is
not too different from $\Mp$, the spectral index becomes close to
$-2\epsilon_\ast$ when this sum is of order $10^3$. This is the
familiar assisted-inflation mechanism. In contrast, the bispectrum
amplitude $\fnl$ has the approximate behaviour
\begin{equation}
	\frac{6}{5} \fnl \approx \frac{2\pi^2}{\Nh}
	\left( \frac{\Mp}{f} \right)^2 ,
	\label{eq:fnlapprox}
\end{equation}
and is independent of $\bar{\epsilon}$ if the dominant fields are
sufficiently close to the hilltop.  It is the different scalings of
$n-1$ and $\fnl$ with $\Nh$ which makes the scenario viable: the
N-flation mechanism lifts the single-field consistency condition $\fnl
\approx - (5/12)(n-1)$ \cite{Maldacena}, which prevents single-field
models generating large non-gaussianity without violating
observational bounds on $n$.

A similar analysis applies to the trispectrum, for which it is
conventional to parameterize the amplitude of a local-type trispectrum
using the parameters $\taunl$ and $\gnl$ of
Eqs.~\eqref{eq:taunl}--\eqref{eq:gnl} \cite{trispectrum},
\begin{eqnarray}
\taunl 
     	&\approx& \left(\frac{4\pi^4}{\Nh^2}\right)
\left(\frac{\Mp}{f}\right)^4 
  	   \approx   \left(\frac{6}{5}\fnl\right)^2 \,,
\label{eq:taunlapprox}\\
\left(\frac{54}{25}\right) \gnl 
  	 &\approx& \left(
\frac{8\pi^4}{\Nh^2}\right)\left(\frac{\Mp}{f}\right)^4 \,.  
\label{eq:gnlapprox}
\end{eqnarray}

Where the summations in Eqs.~\eqref{eq:P}--\eqref{eq:fnl} are
dominated by a single field, this formula shows that the non-gaussian
parameters can become rather large, scaling as powers of $(\Mp/f)^2$.
For $f = \Mp$, we find $\fnl \lesssim 16.4$.  A non-gaussian fraction
of this magnitude should be visible to the Planck satellite.  The same
parameter choice yields $\taunl \lesssim 390$ and $\gnl \lesssim 360$.
Such a small $\gnl$ is unlikely to be observable, although there is
some hope that $\taunl$ of this order could be detected with a future
microwave background polarization satellite \cite{trispec}.

For smaller $f$ the density perturbation becomes increasingly
non-gaussian.  It is even possible to achieve $\fnl \sim 100$ for $f
\sim 0.4 \Mp$, although then $N_{\rm f}$ must be very large to gain
sufficient $e$-foldings and some tension with
Eq.~\eqref{eq:quantum-diffusion-bound} may emerge.  At this $f$, the
trispectrum parameters may become as large as $\taunl \sim 1.5 \times
10^4$ and $\gnl \sim 1.4 \times 10^4$.  Eq.~\eqref{eq:taunlapprox}
shows that when a few hilltop fields dominate the density
perturbation, the axion N-flation model reproduces the single-field
relation between $\fnl$ and $\taunl$ first pointed out by Suyama and
Yamaguchi~\cite{SY}.

\begin{figure}[t]
\includegraphics[width=6 cm,angle=90]{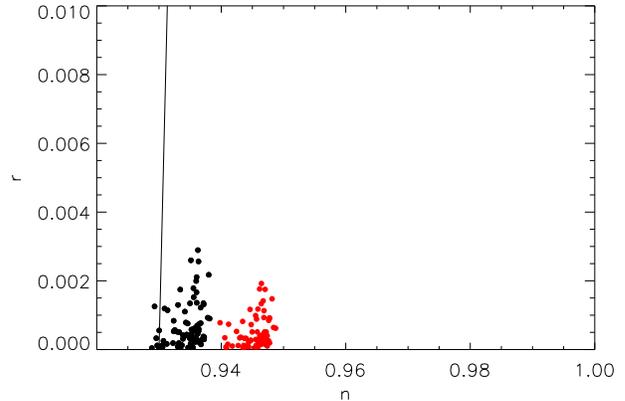}\\
\caption{Predictions in the $n$--$r$ plane, averaged over
  realizations, for various values of $f$ between $0.4\Mp$ and $2\Mp$
  and of $\Nf$ between 464 and 10,000, all giving sufficient
  inflation. The black (left) cluster of points takes $N_* = 50$ and
  the red (right) cluster $N_*=60$. The quadratic expansion predicts
  $r=8/N_*$, far off the top of this plot.  The region right of the
  line is within the WMAP7+BAO+$H_0$ 95\% confidence
  contour~\cite{WMAP7}.
  \label{f:nr}}
\end{figure}

\subsection{Numerical calculations}

To fully explore the model space requires numerical calculations of
the evolution, which we carry out using an extension of the code
developed in Ref.~\cite{KLs}. For each choice of model parameters, a
set of runs is required to explore the uncertainty induced by the
random initial conditions.

In Fig.~\ref{f:nr} we show model predictions in the $n$--$r$ plane,
averaged over several realizations of the initial conditions. We see
$n$ and $r$ are only weakly dependent on the model parameters (though
there is significant dispersion amongst realizations, not shown here),
with the choice of $N_*$ being the principal determinant of $n$.  The
models are compatible with current observational constraints in the
$n$--$r$ plane.

\begin{figure}[t]
\includegraphics[width=6 cm,angle=90]{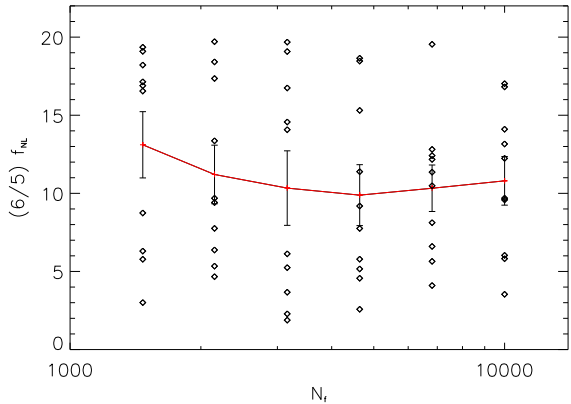}\\
\caption{Predicted non-gaussianity, $\frac{6}{5}\fnl$, for $f=\Mp$ and
  $N_*=50$. The error bars are on the mean over realizations (not the
  standard deviation). Here the maximum achievable value of
  $\frac{6}{5} \fnl$ is $2\pi^2 \simeq 20$, almost saturated in some
  realizations. The significant spread is due to initial condition
  randomness with typical mean values being around half the maximum
  achievable value, and no discernible trend with
  $\Nf$. \label{f:fnl}}
\end{figure}

Turning to the non-gaussianity, Fig.~\ref{f:fnl} shows $\fnl$ as a
function of $\Nf$ for $f = \Mp$, with ten realizations at each $N_{\rm
  f}$. This clearly shows the expected maximum, which is nearly
saturated in cases where a single field dominates the summations. In
cases where several fields contribute significantly to the sums in
Eqs.~\eqref{eq:P}--\eqref{eq:fnl}, the non-gaussian fraction is
reduced.

\begin{figure}[t]
\includegraphics[width=6 cm,angle=90]{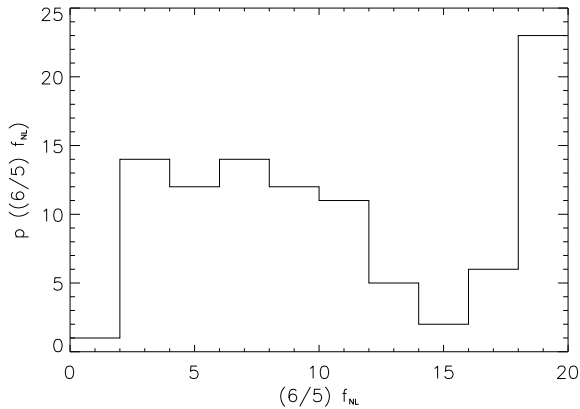}\\
\caption{The distribution of non-gaussianity, $(6/5)\fnl$ due to
  initial condition randomness, for $f=\Mp$ and
  $N_*=50$. \label{f:p_fnl}}
\end{figure}

It is clear from Fig.~\ref{f:fnl} that the non-gaussianity has a large
variance between different realizations of the initial conditions. To
study this in more detail, Fig.~\ref{f:p_fnl} shows the distribution
of $\fnl$ for $f = \Mp$ and $N_* = 50$ with $\Nf = 2150$, now for 100
initial condition realizations. Values near the maximum, corresponding
to the non-gaussianity signal being dominated by a single field, occur
about 25\% of the time, and then there is a broad peak at smaller
values indicating an effective number of contributing fields around
two or more. The broad distribution implies that a measurement of
$\fnl$ alone could not accurately constrain model parameters.

\begin{figure}[t]
\includegraphics[width=6 cm, angle=90]{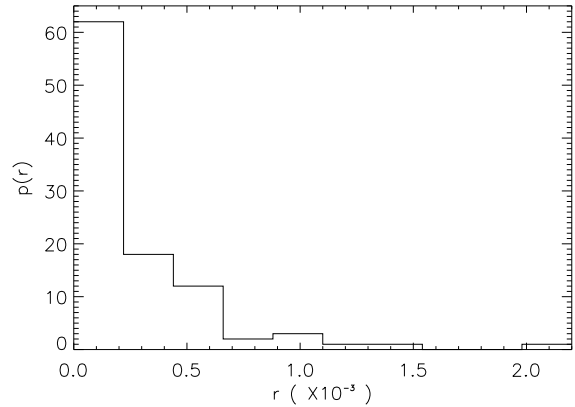}\\
\caption{The distribution of the tensor-to-scalar ratio, $r$, due to
  initial condition randomness, for $f=\Mp$ and
  $N_*=50$. \label{f:p_r}}
\end{figure}

Fig.~\ref{f:nr} shows that we always obtain $r$ values much smaller
than the $8/N_*$ predicted in single-field models. Again there is
significant dispersion from initial conditions, shown in
Fig.~\ref{f:p_r}, with the actual values making any observation
challenging in the extreme. Note that Fig.~\ref{f:p_r} is with the
same conditions as Fig.~\ref{f:p_fnl}.

\begin{figure}[t]
\includegraphics[width=6 cm,angle=90]{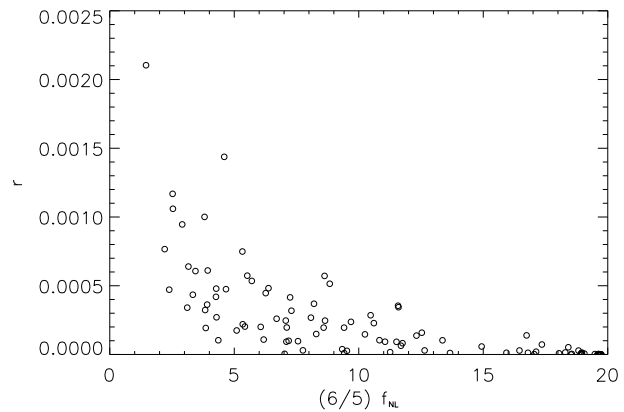}\\
\caption{A scatter plot between the non-gaussianity and the
  tensor-to-scalar ratio, $\fnl$--$r$ due to initial condition
  randomness, for $f=\Mp$ and $N_*=50$. \label{f:fnlr}}
\end{figure}


Moreover, $\fnl$ and $r$ are strongly correlated, as shown in
Fig.~\ref{f:fnlr}, with the larger values of $\fnl$ corresponding to
smaller ones of $r$. The interpretation is that large $\fnl$ requires
one of the fields to be very near the maximum, where the flat
potential forces down the normalization $\Lambda_i$ which then takes
$r$ down as well.

\begin{figure}[t]
\includegraphics[width=6 cm,angle=90]{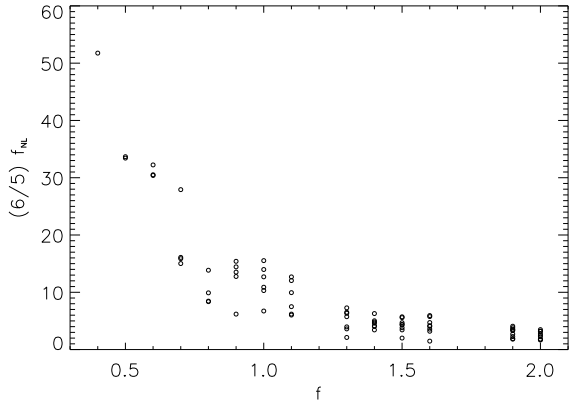}\\
\caption{The predicted non-gaussianity as a function of $f$, for 
  various $\Nf$.\label{f:fnl2}}
\end{figure}

In Fig.~\ref{f:fnl2}, we show the predicted non-gaussianity as a
function of $f$, for a range of choices of $\Nf$. Each point shown is
the average of five or more realizations for an \mbox{$f$--$N_{\rm f}$}
pair. We see a strong trend with $f$, well represented by
Eq.~(\ref{eq:fnlapprox}) with $\Nh \simeq 2$. The different $N_{\rm
  f}$ are scattered by randomness in the initial conditions rather
than an identifiable trend.

\begin{figure}[t]
\includegraphics[width=6 cm,angle=90]{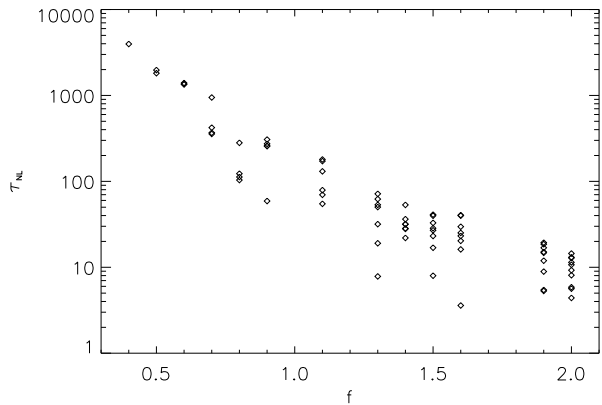}\\
\caption{The predicted trispectrum $\taunl$ as a function
  of $f$, for various $\Nf$. \label{f:ftaunl}}
\end{figure}

Fig.~\ref{f:ftaunl} shows the prediction of the trispectrum of
non-gaussianity $\taunl$, as a function of $f$, for a range of choices
of $\Nf$ as well as ones in $\fnl$. Again each point shown is the
average of five or more realizations for an $f$--$\Nf$ pair, and again
we see the strong trend with $f$, well matched by
Eq.~(\ref{eq:taunlapprox}) with $\Nh \simeq 2$. We have not plotted
the corresponding figure for $(54/25) \gnl$ because it is so similar
to $\taunl$ that it would be almost identical to
Fig.~\ref{f:ftaunl}. This also follows the trend of
Eq.~(\ref{eq:gnlapprox}).

\begin{figure}[t]
\includegraphics[width=6 cm, angle=90]{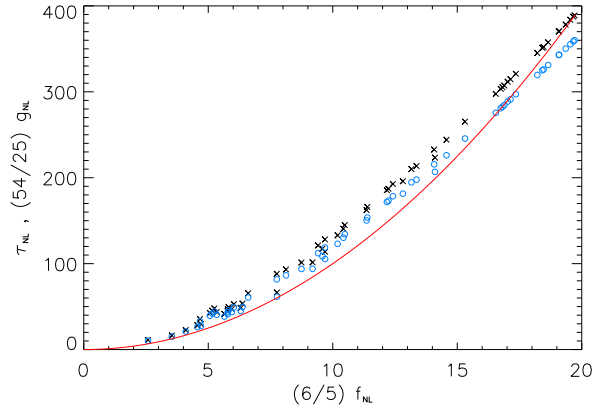}\\
\caption{A scatterplot of the bispectrum $\fnl$ versus the trispectra
  due to initial condition randomness, for $f=\Mp$ and $N_*=50$. Black
  crosses (x) and blue circles (o) are for $\tau_{\rm NL}$ and
  $(54/25)\gnl$ respectively. The red line shows the single-field relation between $\fnl$ and $\tau_{\rm NL}$ from Ref.~\cite{SY}. \label{f:fnltaug}}
\end{figure}

To study the relation between bispectrum and trispectrum,
Fig.~\ref{f:fnltaug} shows the predicted trispectrum as a function of
$\fnl$ for $f=\Mp$ and $N_* = 50$ with $\Nf=2150$. The relation shows
much less scatter than do the individual quantities, due to their
origin in common dynamics. For large $\fnl$, the scatter is at its
smallest, because one field needs to dominate in this case and this
field generates each non-gaussianity parameter in the same way. Suyama and Yamaguchi
 \cite{SY} demonstrated that single-field models satisfy $\tau_{\rm NL} = [(6/5)\fnl]^2$, and that in more general models this expression gives a lower bound to $\tau_{\rm NL}$ (see also Ref.~\cite{SLZ} for a more general derivation). We see that our models do indeed satisfy this inequality on a case-by-case basis, and approach equality in the limit of the highest achievable non-gaussianity.

%
%
\section{Unequal masses}

We consider now the unequal-mass cases. We take the mass spectrum as
exponentially distributed:
\begin{equation}
m_i^2 \equiv m^2 \exp\left(\frac{i-1}{\sigma}\right) ~~~~~~
 {\rm for}~i = 1, 2, \cdots, N_{\rm f}\,,
\end{equation}
where $m$ is the smallest mass. We studied this mass spectrum in the
quadratic potential case in Ref.~\cite{KLs}. Our main objective in
this section is to constrain $\sigma$, which governs how tightly the
mass spectrum is packed.

Choosing the mass spectrum does not fix the model, because the
potentials depend on two parameters which combine to give the mass. We
consider two extreme possibilities.  One is varying the 
amplitude of the potential of each field, $\Lambda_i$, by giving the same value of the constant decay
$f_i = f$ to each field, and the other is varying the decay constant
$f_i$ while keeping the same amplitude $\Lambda_i = \Lambda$. In the
former case the potentials all have the same period but different
amplitudes, and in the latter the same amplitudes and different
periods.

%
%

\subsection{Varying $\Lambda_i$}

Varying $\Lambda_i$ with fixed $f_i = f$ requires
\begin{equation}
   \Lambda_i^2 = \frac{ f m_i}{2\pi } \,.
\end{equation}
If we adopt this in Eqs.~(\ref{eq:P}) to (\ref{eq:gnl}), then we see
that the effect from the different amplitude acts only on the spectral
index and not on $r$ or the non-gaussianity parameters ($\fnl,
\taunl$, and $\gnl$).

\begin{figure}[t]
\includegraphics[width=6 cm,angle=90]{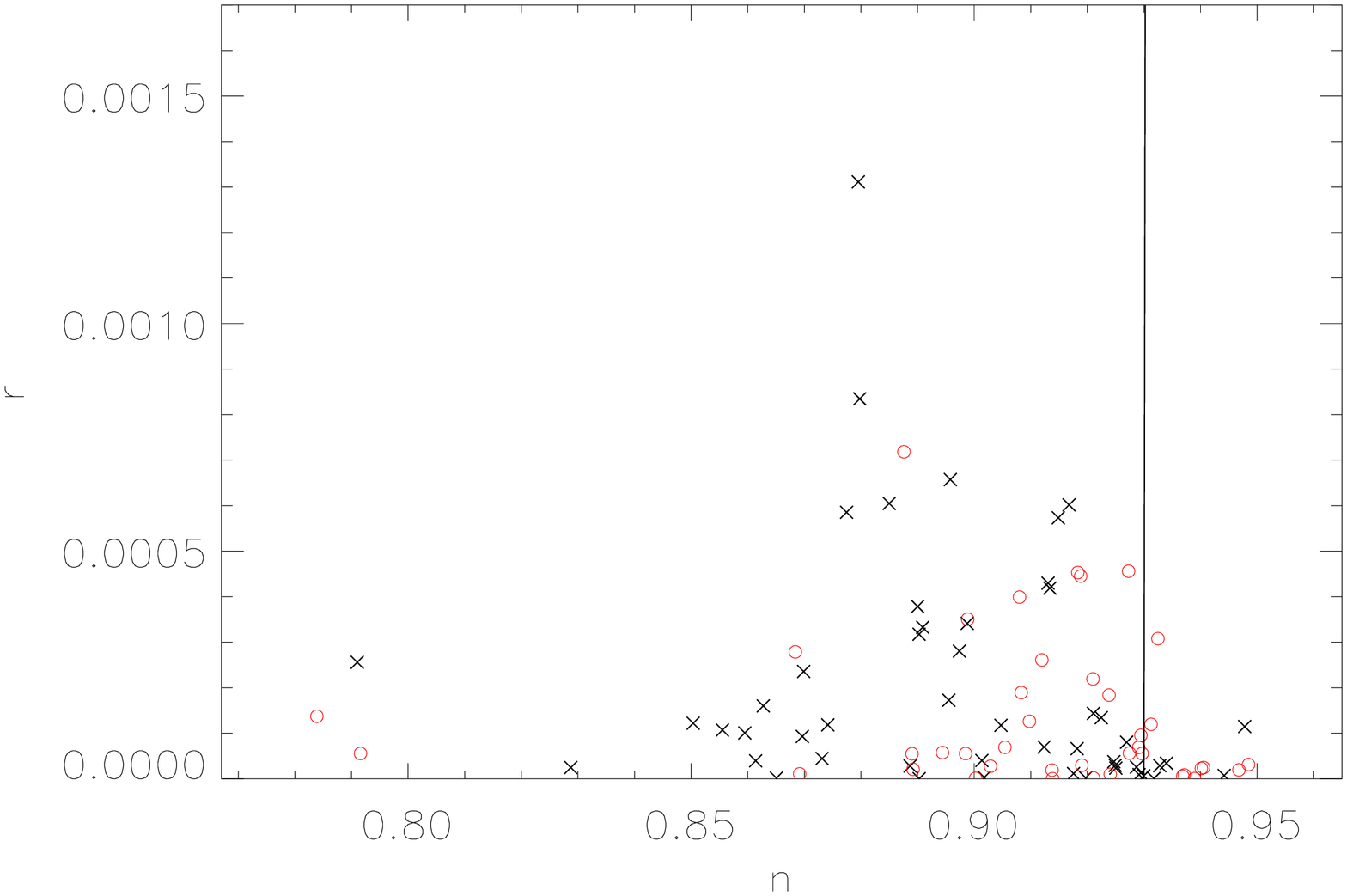}
\includegraphics[width=6 cm,angle=90]{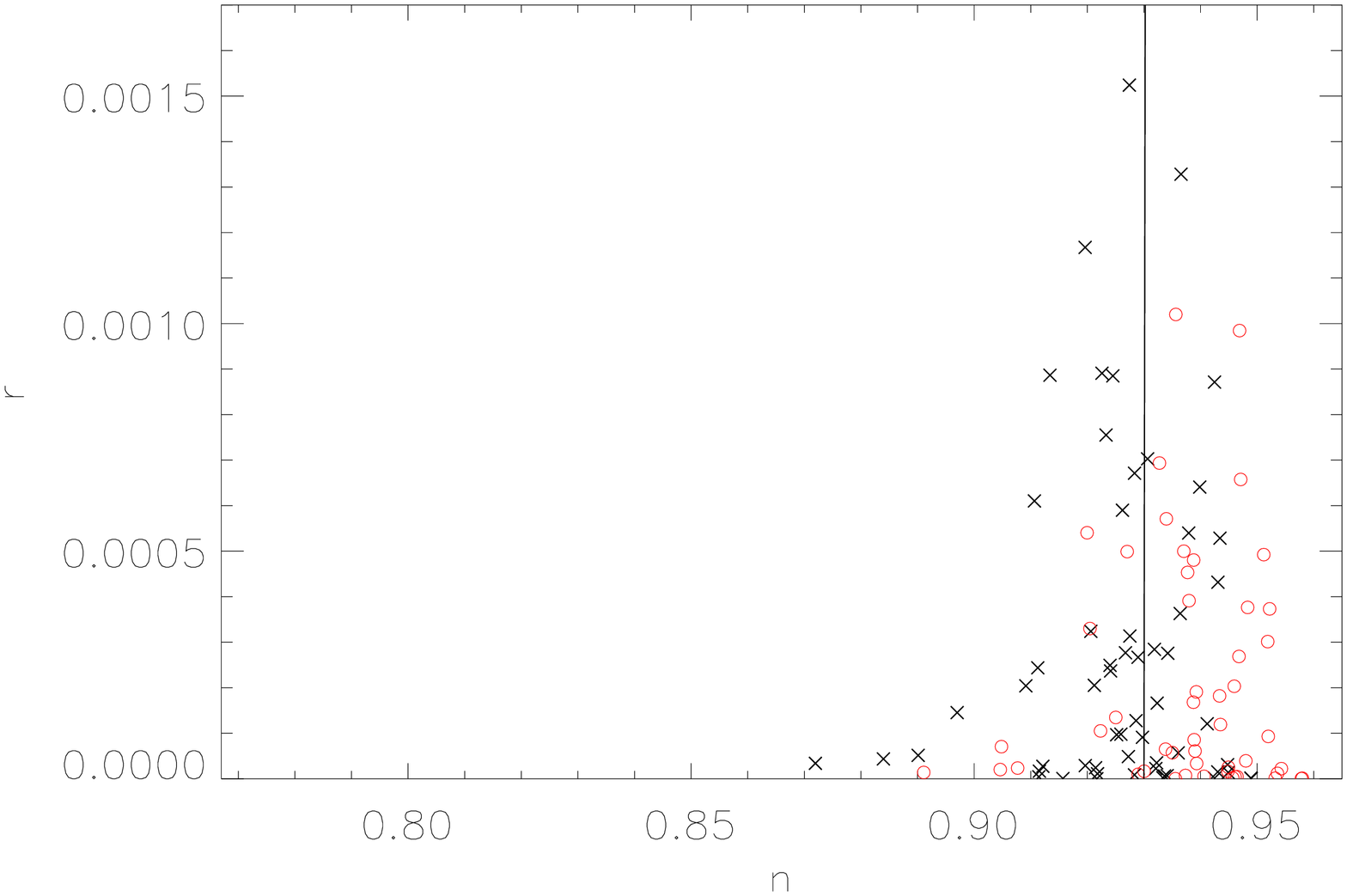}
\includegraphics[width=6 cm,angle=90]{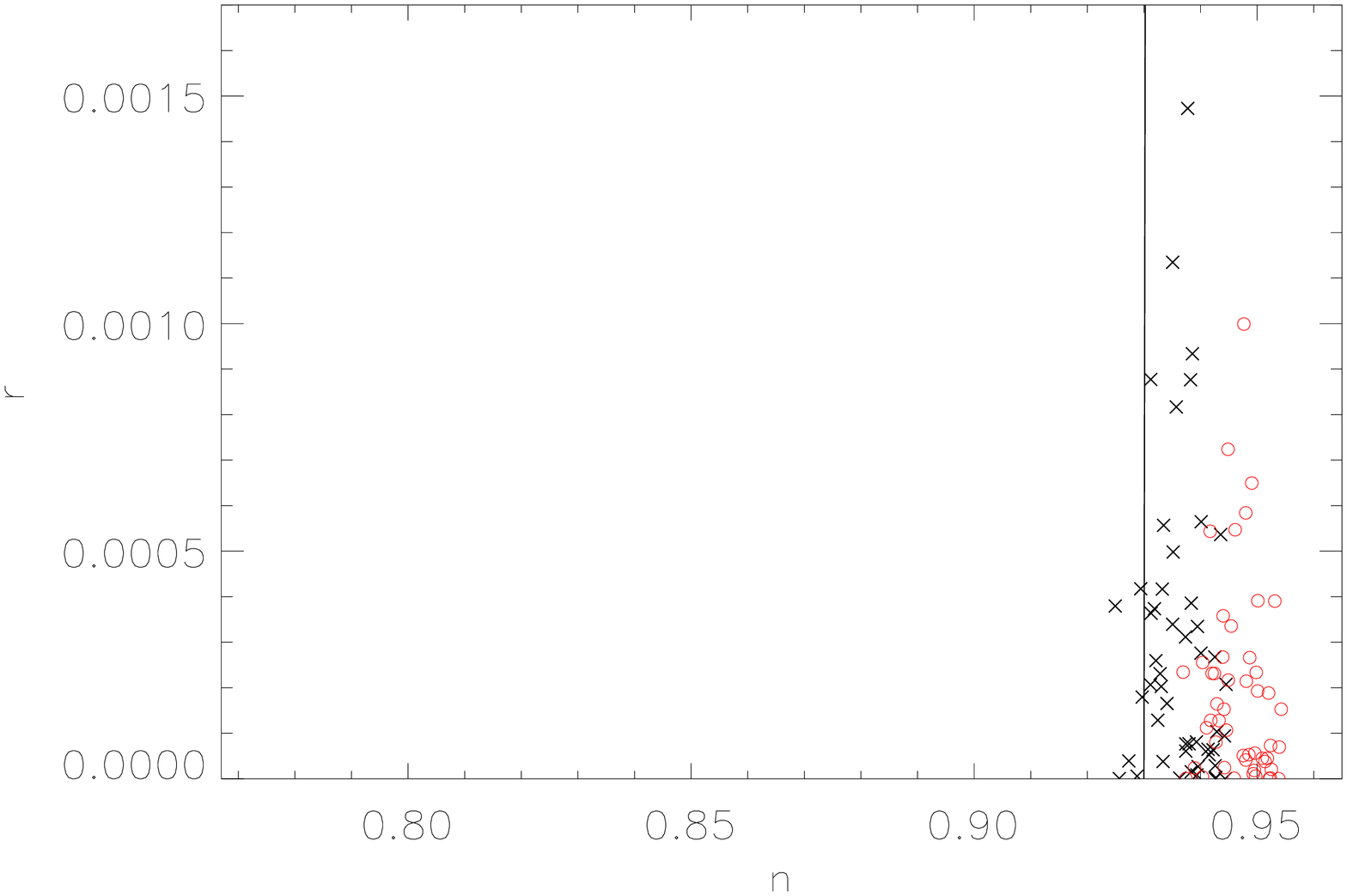}\\
\caption{Predictions in the $n-r$ plane for $\sigma= 500, 2000, 10000$ 
from top to bottom with various $f$ up to a maximum $f = \Mp$. 
The black (left) cluster of crosses takes $N_*=50$ and the red (right) 
cluster of circles $N_* = 60$. Observationally allowed models lie to
the right of the line.
  \label{f:sigmanr2}}
\end{figure}

In Fig.~\ref{f:sigmanr2}, we show the predictions for $n$ and $r$ for
$\sigma=500$, $2000$, and $10000$ with $f = \Mp$. For $\sigma = 500$,
the numerical runs have been done with $\Nf$ in the range 1470 to
6810, for $\sigma =2000$ with $\Nf$ from 1470 to 10000, and for $\sigma = 10000$ case with $\Nf$ from 2150 to 10000. Roughly speaking, for a given $\sigma$ the observables are independent
of $\Nf$. The spectral index
$n$ depends significantly on $\sigma$, but we found the dependence on $f$ is weaker.

\begin{figure}[t]
\includegraphics[width=6 cm,angle=90]{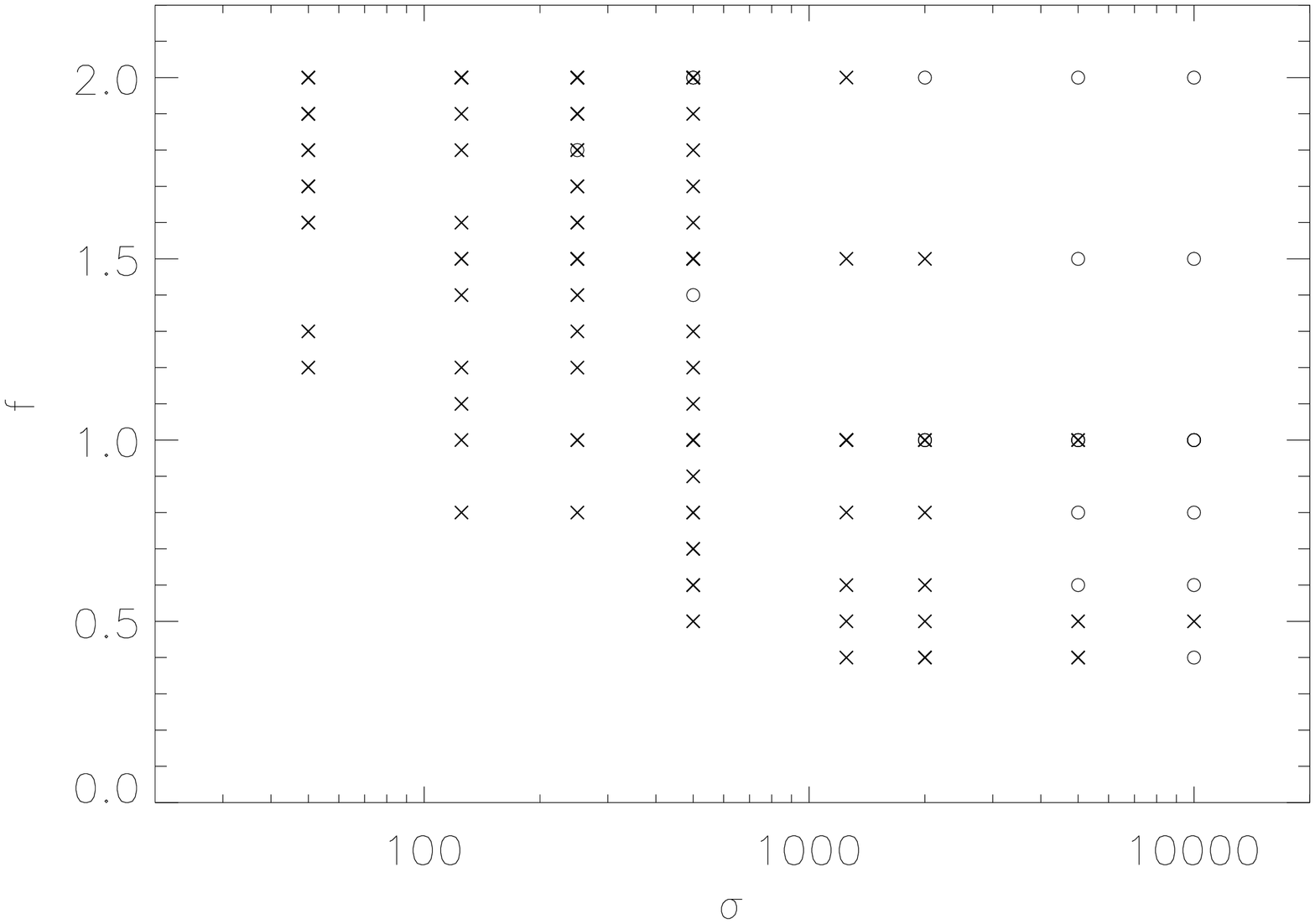}\\
\includegraphics[width=6 cm,angle=90]{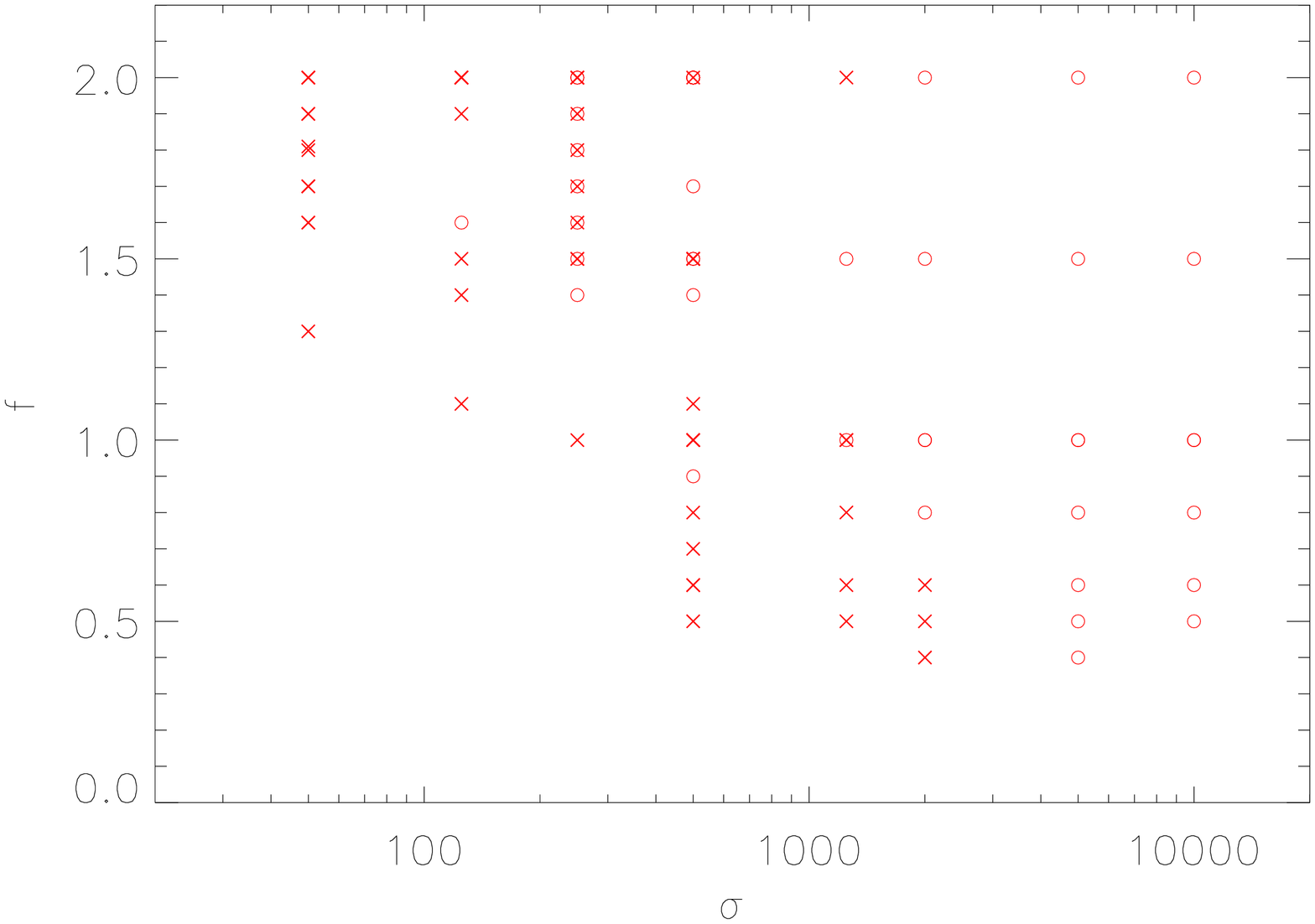}\\
\caption{Locations in the $\sigma$--$f$ plane for $N_*=50$ (top) and
  $N_*=60$ (bottom) where observable predictions for the average value
  $n$ are within the limits (circles, o) or outside them (crosses, x).
\label{f:sigmaf}}
\end{figure}

Small $\sigma$ forces
the spectral index outside of its allowed region in the majority of
cases, leading to a lower limit on $\sigma$.  Looking at various $f$
and $\sigma$, we can map out the allowed parameters.  Figure
\ref{f:sigmaf} samples the $\sigma$--$f$ plane to determine
where the mean value of the spectral index $n$ exceeds the 95\%
observational limit $n=0.93$, for cases with $\Nf$ in the range 464 to 10000. Even though each point is an average over ten initial condition realizations (as many as we could reasonably run), there is still residual noise meaning there is not a perfect partition of the parameter space into allowed and disallowed regions. Nevertheless, the trend is clear; small $\sigma$ is disfavoured, while for large enough $\sigma$ the equal-mass limit is effectively attained which we already know to be viable. Match with data is achieved more comfortably for the larger $N_*$ choice.

The main result from varying $\Lambda_i$ is that the mass spectrum has
to be tightly packed, and then the results from the equal-mass case
are recovered. In
all these cases there is no difference in the non-gaussianities; with
the same $f$ the $\fnl$ does not change.

%
%

\subsection{Varying $f_i$}

\begin{figure}[t]
\includegraphics[width=6 cm,angle=90]{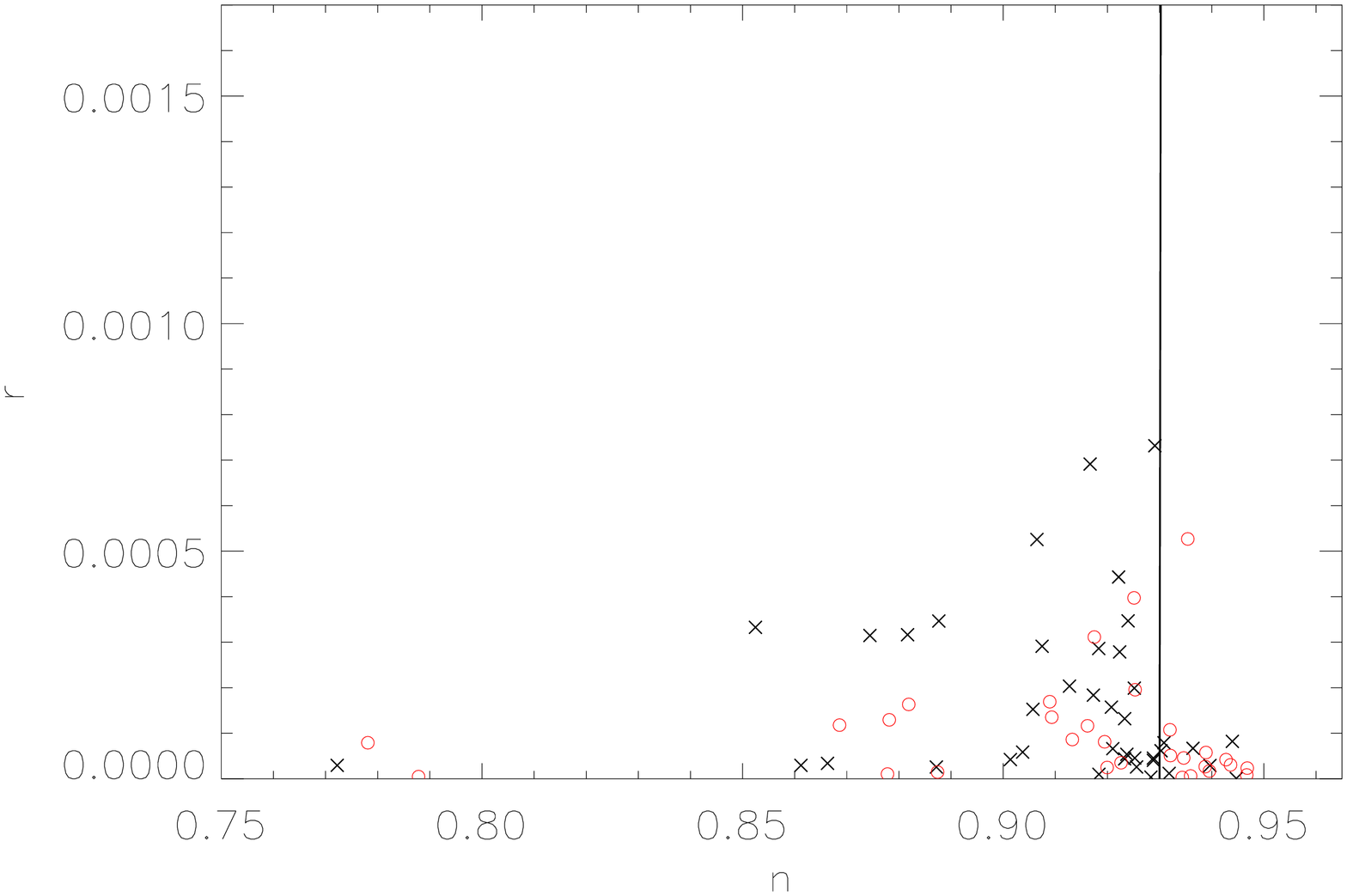}\\
\includegraphics[width=6 cm,angle=90]{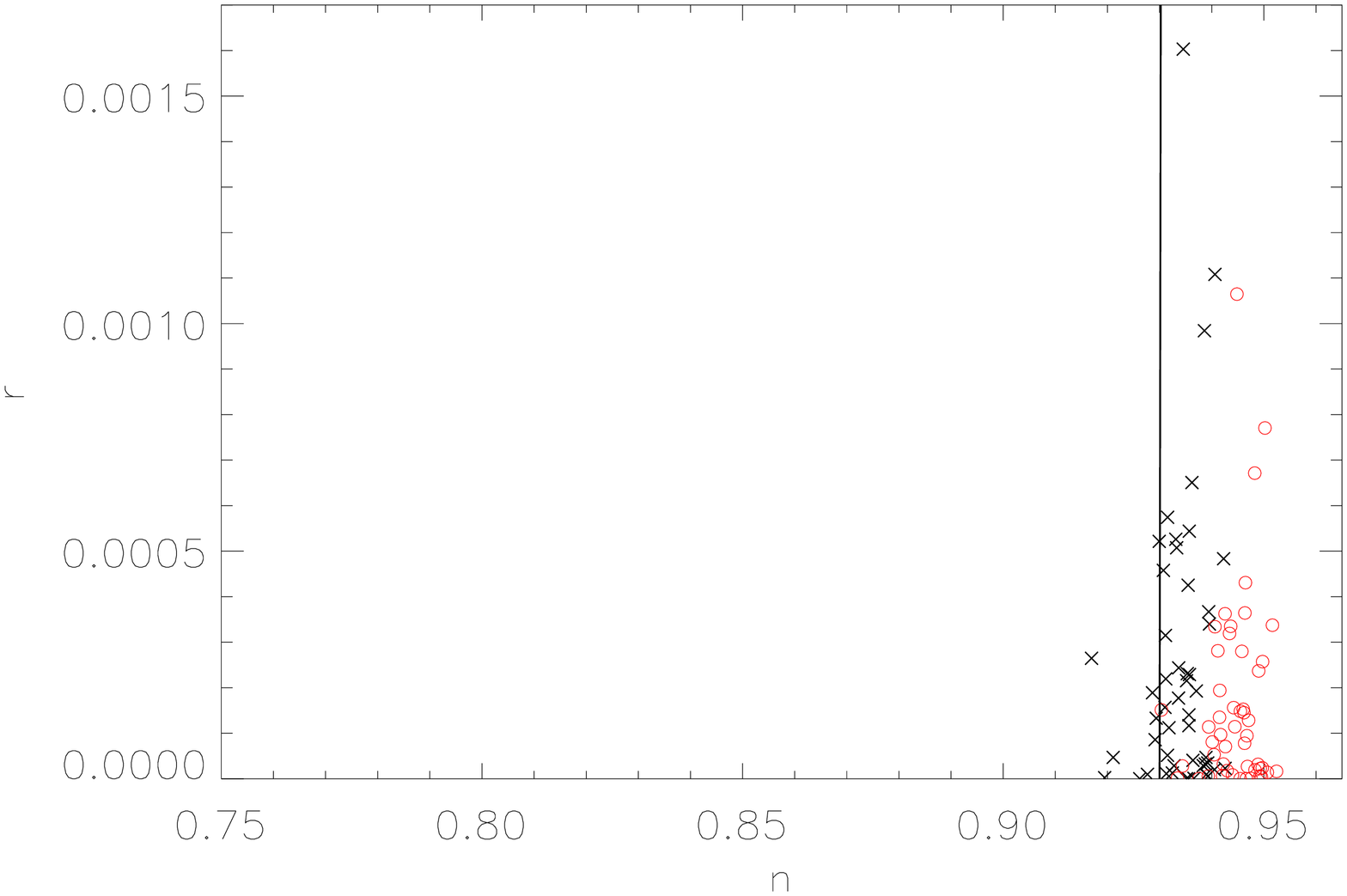}\\
\caption{Predictions in the $n$--$r$ plane for $\sigma= 2000$ (top) and $10000$
  (bottom). The black crosses (x) take
  $N_*=50$ and the red circles (o) $N_* = 60$.
  \label{f:sigmanr3}}
\end{figure}

\begin{figure}[t]
\includegraphics[width=6 cm,angle=90]{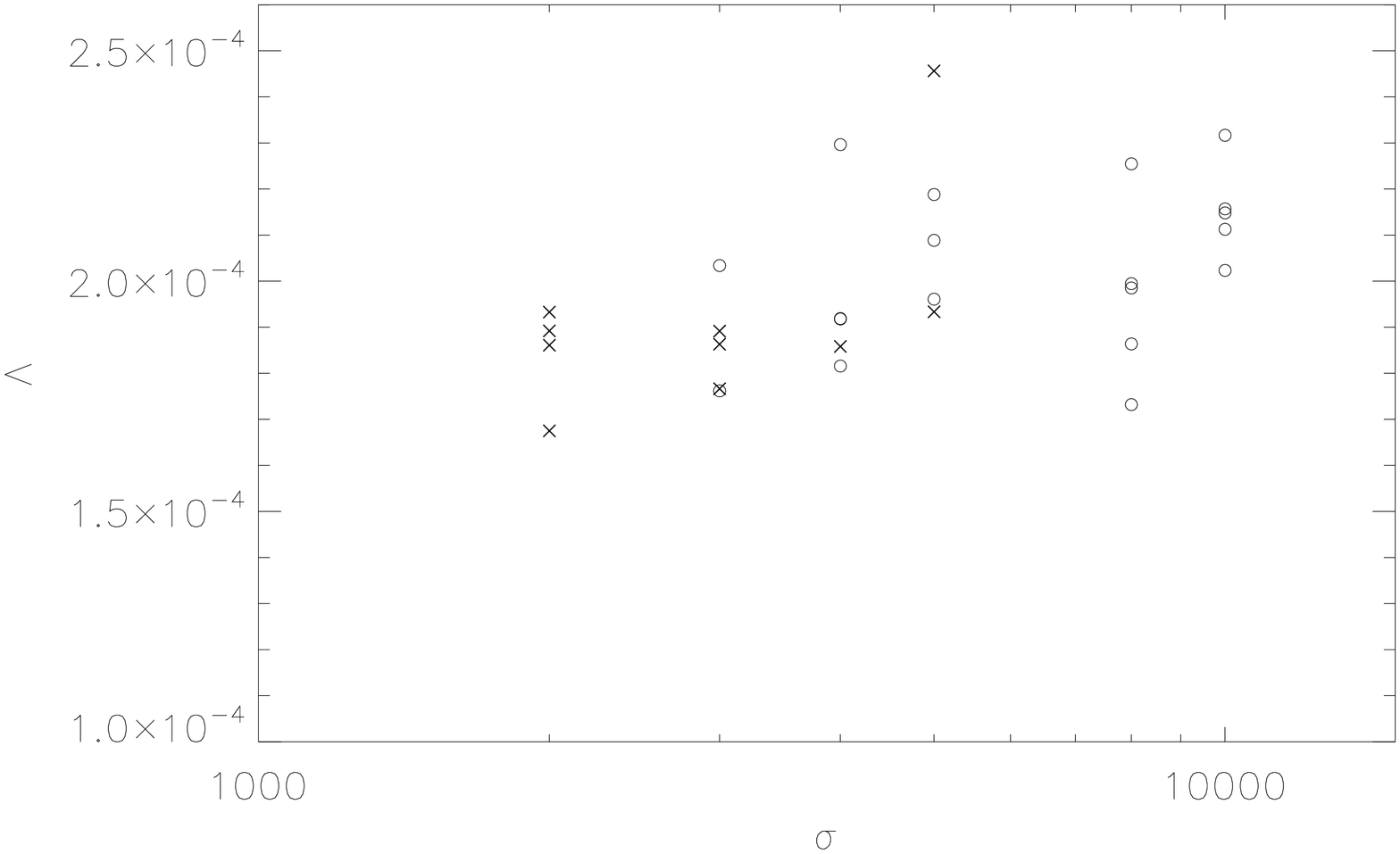}\\
\includegraphics[width=6 cm,angle=90]{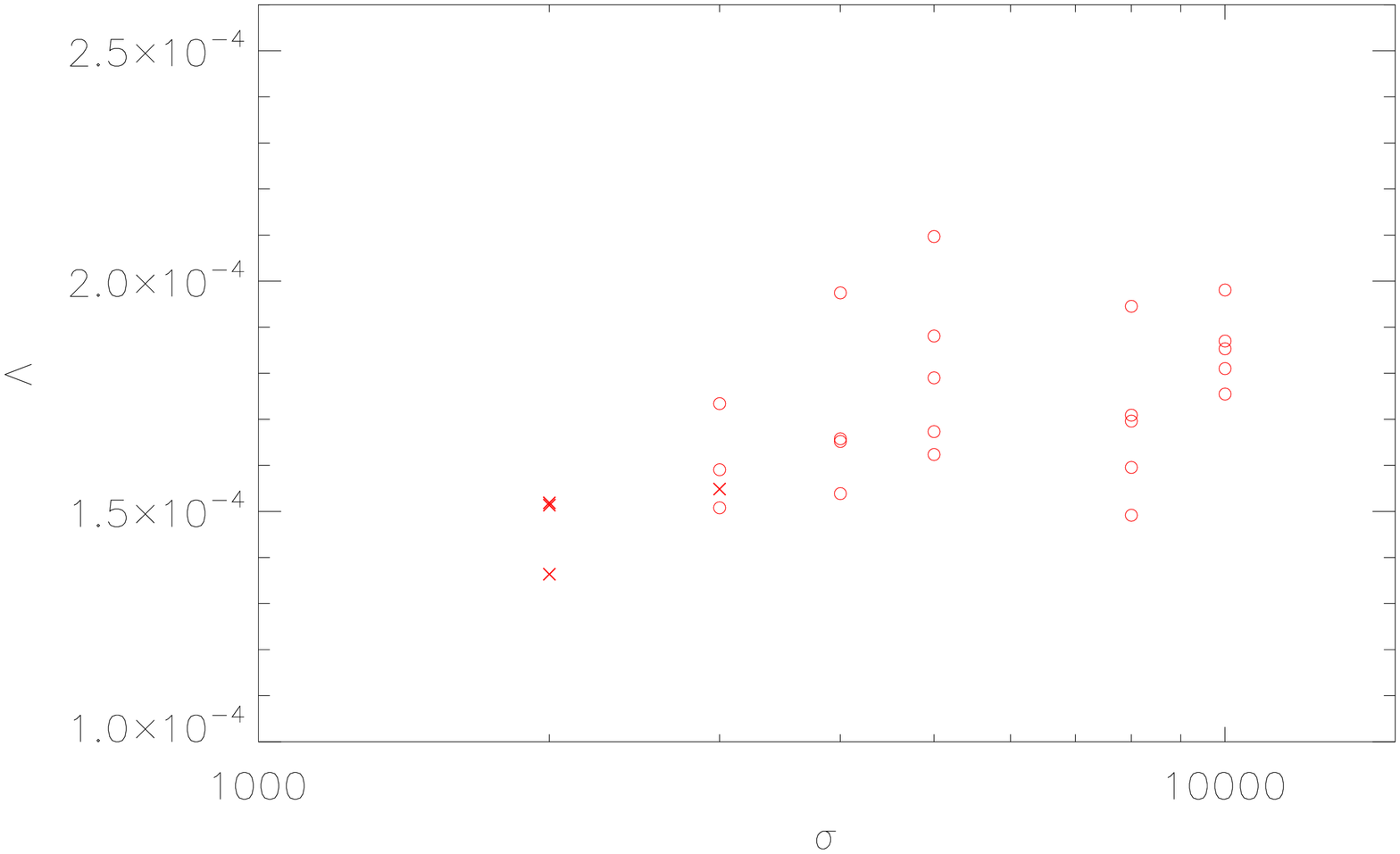}\\
\caption{Predictions in the $\sigma$--$\Lambda$ plane for $N_*=50$
  (top) and $N_*=60$ (bottom), for several choices of $\Nf$. Each point
  shown is the average of ten realizations for an $\sigma$--$\Nf$
  pair. Circles (o) denote that the observable
  predictions for the spectral index are within the limits, and crosses (x) that the values are excluded. \label{f:slambda}}
\end{figure}

The second case fixes $\Lambda_i=\Lambda$, implying
\begin{equation}
f_i = \frac{2\pi \Lambda^2}{m_i} \,,
\end{equation}
where we constrain that the largest $f_i $ is always $\Mp$. Numerical
calculations were done with $\Nf$ in the range 2150 to 10000 and
$\sigma$ from 2000 to 10000. There are not enough $e$-foldings in
cases with $\sigma < 2000$.  The results in the $n$--$r$ plane, and
the regions of parameter space where viable values of those are
achieved, are shown in Figs.~\ref{f:sigmanr3} and \ref{f:slambda}.

The parameter space limits are less clear than in the previous case, as $\Lambda$ is a less fundamental parameter than $f$ whose normalization depends on the particular initial condition realization. Nevertheless, the same general trend is apparent that small $\sigma$ is disfavoured, with no working models found for $\sigma=2000$ regardless of $N_*$. For high enough $\sigma$ the models are always allowed, and in the intermediate regime their validity is a matter for detailed individual analysis.
Once we restrict to observationally-allowed models, the non-gaussianity is essentially that of the equal-mass case though for a smaller `effective' $f$ value given the spectrum of $f_i$ values.


%
%
\section{Conclusions}

We have carried out a detailed study of the phenomenology of the axion
N-flation model, extending our previous work in several
directions. This includes extension of non-gaussianity calculations to
the trispectrum, and an analysis of the correlations between different
observables induced by the initial condition realizations.

When a spectrum of unequal masses is considered, we find that the
spectrum must be extremely tightly packed if the spectral index is to
stay in agreement with observations. This echoes the result we found
for quadratic potentials in Ref.~\cite{KLs}. Once this condition is
obeyed, we find that the predictions for other observable quantities
essentially match those of the equal-mass case, i.e.\ the mass
spectrum does not introduce any new phenomenology.

Nevertheless, the model is highly predictive in terms of all the major
perturbation observables, with the spectral index already close to the
observational lower limit and the non-gaussianity detectable across a
significant volume of model parameter space.


\begin{acknowledgments}
S.A.K.\ was supported by the National Research Foundation of Korea
(NRF) grant funded by the Korean government (MEST) (No.
2011-0011083), and A.R.L.\ and D.S.\ by the Science and Technology
Facilities Council [grant numbers ST/F002858/1 and ST/I000976/1].
S.A.K.\ acknowledges the hospitality of IEU, Ewha Women's University,
of Jihn E. Kim and CTP in Seoul National University, and of the
Astronomy Centre, University of Sussex, while this work was being
carried out. A.R.L\ acknowledges the hospitality of the Institute for Astronomy, University of Hawai`i, while this work was being completed.
\end{acknowledgments}


\end{document}